\documentclass[%
 reprint,
superscriptaddress,
 amsmath,amssymb,
 aps,
floatfix,
]{revtex4-2}

\usepackage{graphicx}
\usepackage{dcolumn}
\usepackage{bm}
\usepackage{url}
\usepackage{xcolor}
\usepackage{hyperref}

\begin{document}
\preprint{APS/123-QED}
\date{\today}

\preprint{Draft: eccentric Galactic binaries in LISA}
\title{Efficient time-domain eccentric model for Galactic binaries in LISA}

\author{Shubhanshu Tiwari}
\affiliation{%
Department of Earth and Planetary Sciences, ETH Zurich,\\
Sonneggstrasse 5, 8092 Z\"urich, Switzerland
}
\author{Johan O. A. Robertsson}
\affiliation{%
Department of Earth and Planetary Sciences, ETH Zurich,\\
Sonneggstrasse 5, 8092 Z\"urich, Switzerland
}
\author{A. Gopakumar}
\affiliation{%
Department of Astronomy and Astrophysics,\\
Tata Institute of Fundamental Research, Mumbai 400005, India
}
\author{Michael Ebersold}
\affiliation{%
Physik-Institut, University of Z\"urich,\\
Winterthurerstrasse 190, Z\"urich, 8057, Switzerland
}
\author{Fredrik P. Andersson}
\affiliation{%
Department of Earth and Planetary Sciences, ETH Zurich,\\
Sonneggstrasse 5, 8092 Z\"urich, Switzerland
}
\author{Maria Haney}
\affiliation{%
National Institute for Subatomic Physics (Nikhef),\\
Science Park 105, 1098 XG, Amsterdam, The Netherlands
}
\affiliation{%
Department of Physics and Astronomy, Vrije Universiteit Amsterdam,\\
De Boelelaan 1100, 1081 HZ Amsterdam, The Netherlands
}
\author{Philippe Jetzer}
\affiliation{%
Physik-Institut, University of Z\"urich,\\
Winterthurerstrasse 190, Z\"urich, 8057, Switzerland
}


\date{\today}
\begin{abstract}
Galactic binaries are expected to be the most numerous long-lived sources in
the Laser Interferometer Space Antenna (LISA) data stream and are commonly
modeled as quasi-circular, nearly monochromatic systems. In this work we
present \texttt{eGB-multi}, a fast time-domain eccentric Galactic-binary
waveform package for LISA. The package combines a post-Newtonian-accurate 
quasi-Keplerian eccentric source model with a time-domain LISA response,
evaluating the signal on the retarded one-way links and then constructing
time-delay interferometry observables. 
Its modular implementation allows the user to switch between different levels of orbital description, ranging from Newtonian closed eccentric orbits and relativistically precessing eccentric orbits incorporating the 1PN-accurate periastron advance, to precessing and shrinking eccentric orbits that additionally account for gravitational radiation reaction at leading quadrupolar order. The framework also provides an automatic evolution mode, in which the appropriate level of orbital dynamics is selected based on a pre-defined mismatch tolerance. We employ this framework to quantify the regimes in which orbital eccentricity renders the commonly adopted quasi-circular approximation for Galactic binaries inadequate.

Using our \texttt{eGB-multi} package, we find that, for detached white-dwarf-like Galactic binaries, the dominant departure from the commonly adopted quasi-circular approximation arises from the richer harmonic structure induced by orbital eccentricity itself.
The periastron advance produces
observable phase and envelope modulations, but remains subdominant for the
wide binaries considered here. 
The quadrupolar order gravitational wave emission 
mainly affects the
accumulated orbital phase at the high-frequency end of the mHz band, while
eccentricity decay is negligible on LISA observation timescales. At the
response level, using exact retarded time-domain links, second-generation
Michelson TDI, and the \(A,E\) channels, we find that quasi-circular templates
lose significant match once the eccentricity becomes large enough, with the
threshold depending primarily on frequency. For a representative mismatch of \(1-\mathrm{FF}=0.1\) which is equivalent to 10\% loss in SNR, quasi-circular
templates remain adequate only below \(e_0\simeq0.07\) at
\(10^{-4}\,\mathrm{Hz}\), with the threshold rising to \(e_0\simeq0.30\) at
\(3\times10^{-3}\,\mathrm{Hz}\). These results motivate eccentric
time-domain response models as practical tools for the detection and parameter
estimation of eccentric Galactic binaries in realistic LISA analyzes.
\end{abstract}


\maketitle

\section{\label{sec:introduction}Introduction}

The Galactic population of compact binaries will form both a dense foreground
and a large catalog of individually resolvable sources for LISA
\cite{AmaroSeoane2017,RobsonCornishLiu2019}. The standard description of
detached Galactic binaries in LISA data analysis treats the sources as
quasi-circular and nearly monochromatic over the mission lifetime. This
approximation is natural for many double white-dwarf binaries, but it need not
hold when a measurable eccentricity is present. Eccentricity redistributes
gravitational-wave power from a single dominant harmonic into a ladder of
orbital harmonics \cite{PetersMathews1963,Peters1964}, and changes the relative
phase and amplitude evolution of the two polarizations. Since the eccentric waveform structure is qualitatively different from the
quasi-circular one, even modest residual eccentricity can require a different
template description. Therefore, if astrophysical formation scenarios populate
the LISA band with eccentric Galactic binaries, an efficient eccentric
Galactic-binary model is needed both to assess the validity of circular
templates and to extract the additional information carried by eccentricity.

Several astrophysical channels motivate the case for eccentric Galactic Binaries. Dynamical interactions in dense stellar systems can produce
eccentric double white dwarfs, unlike the nearly circular disk population
expected from isolated binary evolution \cite{Willems2007}. Recent cluster
studies similarly emphasize eccentric, tight double white dwarfs as a
dynamical subpopulation whose gravitational-wave signals may be useful for
identifying their environments \cite{Hellstrom2025}. Triple evolution provides
another route to LISA-band double white dwarfs and can produce a small number
of systems with extreme eccentricities \cite{Rajamuthukumar2025}. In addition,
Galactic neutron-star--white-dwarf binaries are expected to include eccentric
systems because natal kicks and binary-interaction histories leave imprints on
their eccentricity distributions \cite{Korol2024NSWD}; eccentricity has also
been proposed as a key observable for identifying neutron-star components in
the Galactic-binary population \cite{Moore2024EccentricNS}.

The goal of this paper is to present \texttt{eGB-multi}, a fast and modular
time-domain eccentric Galactic-binary waveform package for LISA, and to use it
to quantify the impact of eccentricity on circular-template searches. The
package combines quasi-Keplerian eccentric source dynamics with a time-domain
LISA response: the source is evaluated on the retarded one-way links and the
resulting link observables are combined into TDI channels. Its source model is
designed to be accuracy-adaptive. 
The user can choose among three levels of dynamical description for Galactic binaries: (i) Newtonian closed eccentric orbits, (ii) relativistically precessing eccentric orbits incorporating the first Post Newtonian-accurate periastron advance, and (iii) precessing and shrinking eccentric orbits that additionally include the leading-order, quadrupolar gravitational-radiation-reaction effects. Most importantly, our ready-to-use pipeline also provides an automatic evolution mode, in which the appropriate level of orbital dynamics is selected based on a user-specified waveform-mismatch tolerance.
This allows the user to easily identify which physical effects control
the mismatch, residuals, and fitting factors without forcing every calculation
to use the most expensive waveform. With this framework we can compute when eccentric signals can still be represented/approximated by
quasi-circular templates, whether circular parameters such as \(f_0\) and
\(\dot f\) can absorb the eccentric structure, and how the computational cost
compares with existing fast circular Galactic-binary response models.

The paper is organized as follows.
Section~\ref{sec:model} describes the eccentric source model, including the
quasi-Keplerian dynamics, harmonic content, periastron advance,
Peters--Mathews evolution, and the source-level implementation in
\texttt{eGB-multi}. Section~\ref{sec:lisa_response} describes the time-domain
LISA response, TDI projection, noise weighting, fitting factors, and the
automatic source-evolution prescription. Section~\ref{sec:results} presents
the response-level circular-vs-eccentric mismatch study and compares the
\texttt{eGB-multi} response with fast circular Galactic-binary response models.
Section~\ref{sec:conclusion} summarizes the implications for eccentric
Galactic-binary searches and future extensions of the package.

\section{Accurate \& Efficient $h_{\times,+}(t)$ for Galactic Binaries in General Relativistic 
Eccentric orbits}{\label{sec:model}
Our approach to compute $h_{\times,+}(t)$ from non-spinning comparable mass galactic binaries 
in precessing eccentric orbits is influenced by Refs.~\cite{TessmerGopakumar2006,TanayHaneyGopakumar2016,DGI}.
Therefore, we begin by listing the quadrupolar order $h_{\times,+}$  due to non-spinning compact binaries in non-circular orbits
\begin{align}
h_+(t) &= -\frac{G M \eta}{c^4 D}
\Bigl[
(1+\cos^2\iota)
\left(a\cos 2\phi+b\sin 2\phi\right)
\nonumber\\
&\hspace{2.6cm}
+\sin^2\iota\,c_r
\Bigr], \\
h_\times(t) &= -\frac{2G M \eta}{c^4 D}\cos\iota
\left(a\sin 2\phi-b\cos 2\phi\right),
\end{align}
where the total binary mass $M = m_1 +m_2$ while $\eta$ stands for the 
symmetric mass ratio (
$\eta=m_1m_2/M^2$).
Further, $\iota, D $ denote the binary's orbital inclination and the luminosity distance, respectively while the coefficients $a,b$ and $c_r$ are functions of binary's dynamical variables.  Influenced by Ref.~\cite{TessmerGopakumar2006}, we write 
\begin{align}
a &= \frac{GM}{r}+r^2\dot\phi^2-\dot r^2,
\nonumber\\
b &= 2r\dot r\dot\phi,
\nonumber\\
c_r &= \frac{GM}{r}-r^2\dot\phi^2-\dot r^2\,,
\end{align}
where  $r$ and $\phi$ define the components of the relative separation 
$ \vec{r} = r(\cos \phi,\sin \phi, 0)$, and their time derivatives are denoted 
by $\dot r$ and $\dot \phi$.

 To construct temporally evolving $h_{\times,+}(t)$ for galactic binaries in 
 precessing eccentric orbits, we employ the quasi-Keplerian parametric solution \cite{DD85}.
 This  parametrization  extends the classical Keplerian solution to relativistic compact binaries by expressing the post-Newtonian accurate orbital motion in terms of generalized orbital elements and anomalies.
 It provides an accurate and computationally efficient description of eccentric binary dynamics, forming the basis for both GW waveform modeling and precision pulsar timing
 \cite{DGI,DD86}.
Recall that the post-Newtonian (PN) approximation to general relativity allows us to express the equations of motion for a compact binary as corrections to
Newtonian equations of motion in powers of $ (v/c)^2 \sim G\, M/ (c^2\, r) $, 
 where $ v, M$, and $r$ are the
velocity, total mass and relative separation of the binary.
The use the quasi-Keplerian paramteric solution ensures that each galactic binary is 
specified by the following parameters
\begin{equation}
\bm{\theta}_{\rm src} =
\{\omega_0, e_0, m_1, m_2, D, \iota, \phi_0, t_0\},
\end{equation}
where $\omega_0$ is certain sidereal angular frequency at the
reference time $t_0$,and $\phi_0$ the associated initial
orbital phase while 
$e_0$ is essentially the eccentricity parameter of the quasi-Keplerian parametric solution. 

\subsection{\label{subsec:fixed_eccentric} Time-Domain Waveforms for Galactic Binaries 
in Closed/Open Eccentric Orbits}

We begin by briefly summarizing the classical 
 Keplerian parametric solution that describes  Newtonian accurate 
orbital motion of a binary in closed 
eccentric orbits
\begin{align}
r &= a(1-e\cos u )\,,
\\
\phi -\phi_0 &=v\equiv 2 \arctan \biggl [ \biggl ( \frac{ 1 + e}{ 1 - e}
\biggr )^{1/2} \, \tan \frac{u}{2} \biggr ]\,,
\end{align}
where $a$ and $e$ stand for the  semi-major axis and the eccentricity of the orbit, 
respectively while the auxiliary angles $u$ and $v$ are 
called  eccentric and true anomaly, respectively.
The explicit time dependence is provided by the Kepler equation, which reads
\begin{align}
l \equiv n (t - t_0)  &= u - e\,\sin u\,,
\end{align}
where $l$ is the mean anomaly and $n$ is referred to as the mean motion 
and is given by
$ n = 2\,\pi/P$, $P$ being the orbital period.
It is important to note that this parametric solution allows us to express analytically 
$r, \dot r, \phi$ and $\dot \phi$ in terms of $n,e$ and $u$.
Further, the temporal evolution of 
these dynamical variables is obtained by solving the transcendental Kepler equation and imposing the resulting $u(l)$ on the analytic expressions for $r, \dot r, \phi$ and $\dot \phi$.
With the help of Eqs.~ for $h_{\times,+}$, this leads to temporally evolving quadrupolar 
order GW polarization states for galactic binaries in closed (Newtonian) eccentric orbits.

 The quasi-Keplerian parametric solution, detailed in Ref.~\cite{DD85},
 allows us to describe 1PN-accurate trajectory of galactic binary 
 which due to the general relativistic advance of periastron leads to open/precessing 
 eccentric orbits. This 
`Keplerian like' parametrization is given by 
\begin{align}
r &= a_r(1-e_r \cos u)\,,\\
l \equiv n (t -t_0) &= u - e_t \sin u\,,\\
(\phi-\phi_0) &=( 1+k) \, v \,, {\mbox{where}\,\,} 
\nonumber\\
 v & \equiv
2 \arctan \biggl [ \biggl ( \frac{ 1 + e_{\phi}}{ 1 - e_{\phi}}
\biggr )^{1/2} \, \tan \frac{u}{2} \biggr ]\,.
\end{align}
 We would like to note that these three eccentricity 
 parameters $e_r, e_{\phi}$ and $e_t$ were introduced to 
 ensure that the PN-accurate parametrization looks `Keplerian' even at 1PN order 
 and these eccentricities 
 differ from each other by PN corrections.
 The parameter $k$ stands fo the rate of periastron advance per orbit which ensures 
 that trajectory of the underlying compact binary remains an open (or precessing )
 eccentric orbit.
Note that $n \times k$  quantifies the relativistic advance of periastron per orbital revolution. In the case of the Double Pulsar, PSR~J0737$-$3039A/B, this corresponds to an advance of about $(14^{\prime\prime})$  per orbit, or equivalently $(16.9^{\circ}\mathrm{yr}^{-1})$, highlighting the strongly relativistic nature of 
the system \cite{Kramer2021DoublePulsar}.
Likewise, many compact Galactic binaries detectable by LISA will reside in similarly strong-field relativistic regimes, where post-Newtonian effects, including periastron advance, will play a central role in modeling their orbital dynamics and gravitational-wave emission.

 It is now straightforward to compute 1PN-accurate analytic expressions for 
$r, \dot r, \phi$ and $\dot \phi$ in terms of $n,e_t$ and $u$.
Moreover, the temporal evolution of 
these dynamical variables is obtained by solving the 1PN-accurate 
transcendental Kepler equation, namely $l= u -e_t\,\sin u$
and imposing the resulting $u(l)$ on the PN-accurate analytic expressions for $r, \dot r, \phi$ and $\dot \phi$.
This now leads to certain restricted post-Newtonian accurate $h_{\times,+} (t)$
for galactic binaries in precessing eccentric orbits.
The time-domain template family is restricted as PN-accurate orbital evolution is 
imposed on the quadrupolar 
order GW polarization states. 

We now list symbolically 1PN-accurate expressions for dynamical varaibles that appear 
  in Eqs.~ for $h_{\times,+}$
\begin{align}
r &=
\frac{GM}{c^2x}
(1-e_t\cos u)
\left[
1+x\,r_{\rm 1PN}(\eta,e_t,u)
\right],
\nonumber\\
\frac{\dot r}{c} &=
\frac{\sqrt{x}\,e_t\sin u}{1-e_t\cos u}
\left[
1+x\,\dot r_{\rm 1PN}(\eta,e_t)
\right],
\nonumber\\
\dot\phi &=
\frac{c^3}{GM}
\frac{\sqrt{1-e_t^2}\,x^{3/2}}
{(1-e_t\cos u)^2}
\left[
1+x\,\dot\phi_{\rm 1PN}(\eta,e_t,u)
\right],
\end{align}
where the coefficient functions are those of the 1PN part of Eqs.~(3.7) and
Appendix~B of Ref.~\cite{TanayHaneyGopakumar2016}.
Further, we employ the widely used PN expansion parameter 
$ x=\left(\frac{GM\omega}{c^3}\right)^{2/3} $ to obtain PN-accurate expressions for the relevant dynamical variables, where $ \omega $ is the sidereal angular frequency such that 
$ \omega = n ( 1+k)$. 

 Influenced by Ref.~\cite{DGI}, we write the expression for $\phi$ in a form that explicitly separates the secular periastron advance from the periodic orbital motion.
This leads to 
\begin{align}
\phi &= \lambda + W,
\end{align}
where $ \lambda = ( 1+k_{\rm 1PN} ) l$ while the orbital time scale periodic variations are given by 
\begin{align}
W &= (v-u)+e_t\sin u
+x\,W_{\rm 1PN}(\eta,e_t,u)\,.
\end{align}
The explicit 1PN contribution to $W$ is given by Eq. in Ref.~\cite{TanayHaneyGopakumar2016}
Further, we employ the following expressions to evaluate $\phi$:
\begin{align}
k_{\rm 1PN} &=\frac{3x}{1-e_t^2}\,,\\
v-u &=
2\tan^{-1}
\left[
\frac{\beta_\phi\sin u}{1-\beta_\phi\cos u}
\right],
\end{align}
with
\begin{align}
\beta_\phi &=
\frac{1-\sqrt{1-e_t^2}}{e_t}
+x\,\beta_{\phi}^{\rm 1PN}(\eta,e_t).
\end{align}
It is important to note that this prescription for $\phi$ ensures that 
the correct circular limit naturally arises in our approach. In other words, when 
$e_t\rightarrow0$, we have  $W\rightarrow0$ and
$\phi\rightarrow\lambda=\omega(t-t_0)+\lambda_0$ as $ ( 1+k)n \equiv \omega$

  Interestingly, the 1PN-accurate Kepler equation is structurally identical to the classical 
  Kepler equation:
\begin{equation}
l = u-e_t\sin u+\mathcal{O}(x^2)\,,
\end{equation}
where PN corrections appear at the 2PN order and we employ Mikkola's method to solve 
it to obtain $u(l,e_t)$ \cite{Mikkola1987}.
Note that 
the Kepler equation is transcendental in nature and therefore, cannot be solved analytically in terms of elementary functions. The prescription developed by S.~Mikkola employs 
 an analytic approximation to the eccentric anomaly followed, if necessary,
 by at most one Newton--Raphson refinement.
 This approach provides solutions with nearly machine precision over the entire range of bound eccentricities $ (0 \leq e < 1  $) while being computationally efficient and robust, particularly for highly eccentric orbits.
 
In this approach, the binary evolves through
a sequence of quasi-Keplerian orbits whose orbital elements, in particular the
azimuthal angular frequency $\omega$ and the time eccentricity $e_t$, evolve
secularly according to radiation-reaction equations. To understand whether GW-emission-induced secular evolution matters for Galactic
binaries, we adapt the gravitational-wave phasing formalism developed in
\cite{DGI}. We evolve the variables
$\omega$, $e_t$, $l$, and $\lambda$ on the waveform time grid using
\begin{subequations}
\label{Eq_GWphasing}
\begin{align}
\frac{d\omega}{dt}
&=
\frac{96}{5}\eta
\frac{(GM)^{5/3}}{c^5}
\omega^{11/3}
\frac{1+\frac{73}{24}e_t^2+\frac{37}{96}e_t^4}
{(1-e_t^2)^{7/2}},
\label{eq:omegadot} \\
\frac{de_t}{dt}
&=
-\frac{304}{15}\eta
\frac{(GM)^{5/3}}{c^5}
\omega^{8/3}
\frac{e_t\left(1+\frac{121}{304}e_t^2\right)}
{(1-e_t^2)^{5/2}},
\label{eq:edot} \\
\frac{dl}{dt}
&=
n_{\rm 1PN},
\label{eq:ldot}
\\
\frac{d\lambda}{dt}
&=
\omega \,,
\label{eq:lamdot}
\end{align}
\end{subequations}
where $ n_{\rm 1PN}$ stands for 
the 1PN-accurate radial mean motion, and we express it in terms of 
$\omega$. This expression is extractable from the relation $ \omega = n ( 1+k) $ 
and it reads 
\begin{equation}
n_{\rm 1PN}
=
\omega
\left (
1-\frac{3x}{1-e_t^2}
\right ),
\label{eq:n_omega_relation}
\end{equation}
where $x$ as noted earlier is defined to be $
x \equiv  \left(\frac{GM\omega}{c^3}\right)^{2/3}.
$
Note that in Eqs.~\ref{Eq_GWphasing},  the first two equations are the leading quadrupolar radiation-reaction
contributions, formally entering at 2.5PN order
\cite{PetersMathews1963,Peters1964,DGI}. The third equation uses
Eq.~(\ref{eq:n_omega_relation}) to convert the evolved azimuthal frequency
into the radial mean motion that appears in the Kepler equation. This gives
$\omega(t)$, $e_t(t)$, $l(t)$, and $\lambda(t)$, which are then inserted into
the 1PN quasi-Keplerian orbital variables after solving
$l=u-e_t\sin u$ for $u(l,e_t)$ using Mikkola's method. 
Imposing these variations in Eqs.~  for the quadrupolar order GW polarization states, we 
obtain in a computationally efficient way $h_{\times,+}(t)$ associated with inspiraling  Galactic binaries 
in 1PN accurate precessing eccentric orbits.

It is now straightforward to compute the complex source strain convention which combines the two GW polarizations into a single complex quantity. 
 We define the complex source strain as 
\begin{equation}
h(t)=h_+(t)-i h_\times(t).
\end{equation}
This strain is introduced as it is convenient for spin-weighted spherical harmonic decomposition, and detector response calculations.
Interestingly, we may introduce the Fourier series expansion for either 
$+$ or $ \times$ polarization state as 
\begin{align}
h_P(l) &=
\sum_{k=-\infty}^{\infty}
H^P_k(\bm{\theta}_{\rm src})\,e^{i k l},
\nonumber\\
H^P_k &=
\frac{1}{2\pi}\int_0^{2\pi}
h_P(l)\,e^{-i k l}\,dl,
\label{eq:eccentric_harmonic_expansion}
\end{align}
where, $P=+,\times$.
If  we neglect effects GW emission and periastron advance, we have 
$l=n(t-t_0)$  and $k$ is the
integer harmonic index of the radial orbital frequency $n/(2\pi)$. Since
$h_+$ and $h_\times$ are real
time series, $H^P_{-k}=(H^P_k)^*$. For the complex strain convention used here,
\begin{equation}
h(l)=h_+(l)-i h_\times(l)
=\sum_{k=-\infty}^{\infty}
\left(H^+_k-iH^\times_k\right)e^{i k l}.
\end{equation}


Let us now explore the distinct phenomenological features of our 
time-domain eccentric polarization states $h_{\times,+}(t)$.
The most drastic effect is already present in Newtonian order.  For a
circular binary, the quadrupolar waveform is essentially monochromatic, with
the dominant emission concentrated at twice the orbital frequency.  In an
eccentric orbit, however, the orbital separation and angular velocity vary
within each radial period.  The resulting waveform is no longer sinusoidal, this means that the eccentricity redistributes power from the single central frequency into a set
of harmonics of the radial orbital frequency.  This harmonic structure is the
leading source-level reason why eccentric Galactic binaries cannot always be
represented accurately by the usual circular, nearly monochromatic template,
even if the central frequency and a small frequency derivative are adjusted.

\begin{figure}[!htbp]
\includegraphics[width=\columnwidth]{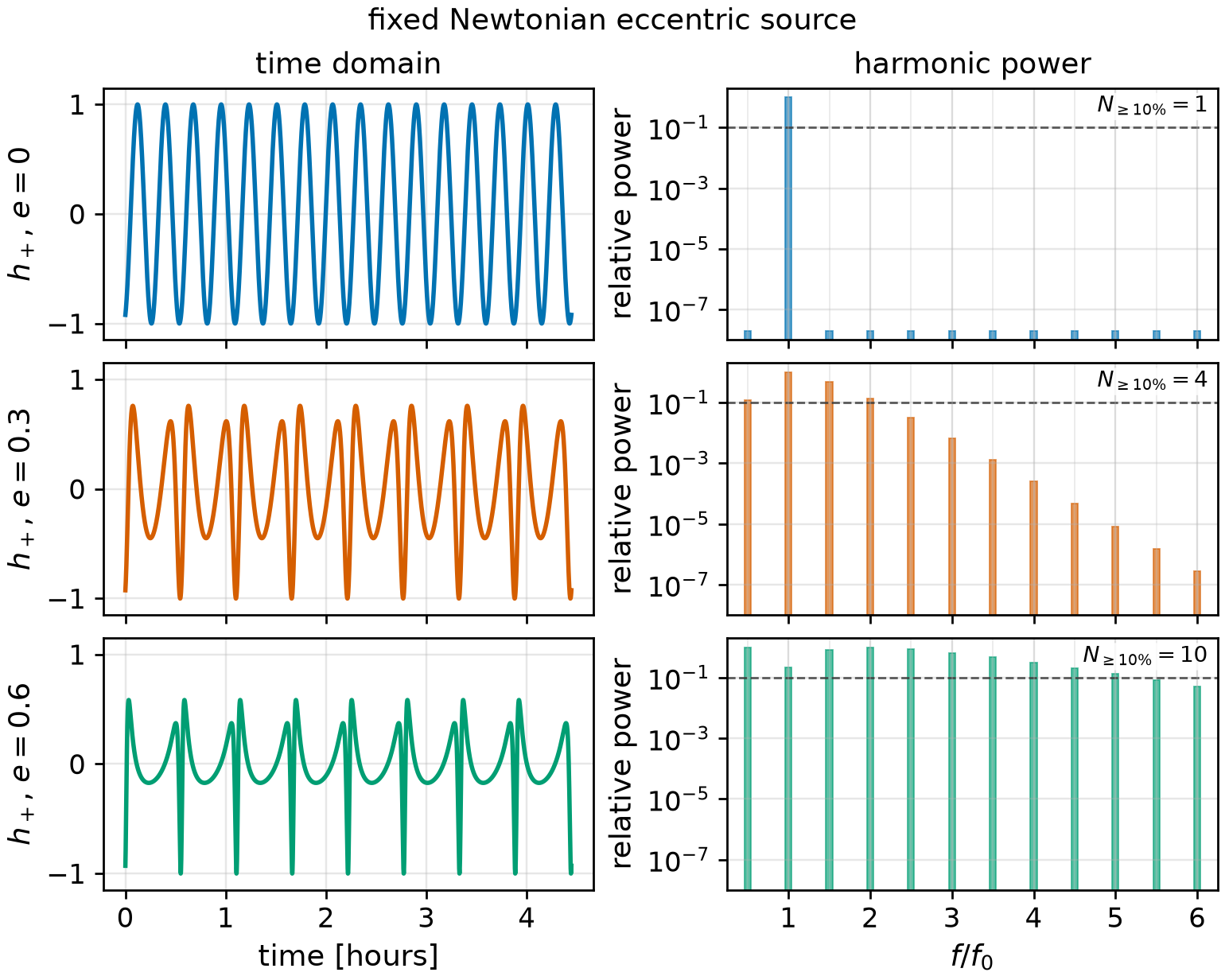}
\caption{\label{fig:source_harmonics}
Time-domain source waveforms and harmonic power for eccentricities
$e=0$, $0.3$, and $0.6$, generated with the same fixed Newtonian
Galactic-binary parameters before applying the LISA response. Each row shows
one eccentricity. Left column: Normalized $h_+(t)$ over the first few hours.
Right column:
Relative harmonic power, normalized to the strongest harmonic in each row and displayed at integer harmonics of the radial orbital frequency, $f/f_0= i/2$, where 
$i$ is an integer, and 
 $f_0$ denotes twice the radial orbital frequency. 
The horizontal dashed line marks the $10\%$ relative-power level,
and $N_{\geq10\%}$ gives the number of harmonics above this threshold. In the
circular limit the signal is concentrated in the $i=2$ harmonic, at
$f/f_0=1$. As eccentricity increases, power is redistributed across a wider
set of harmonics, demonstrating that eccentricity introduces multi-harmonic
structure already at Newtonian order.}
\end{figure}

This behavior is illustrated in Fig.~\ref{fig:source_harmonics}, where the
time-domain polarization visibly departs from a sinusoid as $e_t$ increases
and the Fourier spectrum develops sidebands around the circular carrier.  This 
effect does not rely on radiation reaction, periastron advance, or any higher
relativistic correction; those effects modify the phasing and line positions,
but the appearance of multiple harmonics is a direct consequence of Newtonian
eccentric motion. The present source model is therefore built to provide a
natural extension of the standard monochromatic Galactic-binary model: it
keeps the same physical source parameters in the circular limit, but replaces
the central frequency waveform by a time-domain eccentric polarization that can
be propagated through the same LISA response and TDI analysis chain.

\begin{figure*}[!tbhp]
\includegraphics[scale=0.5]{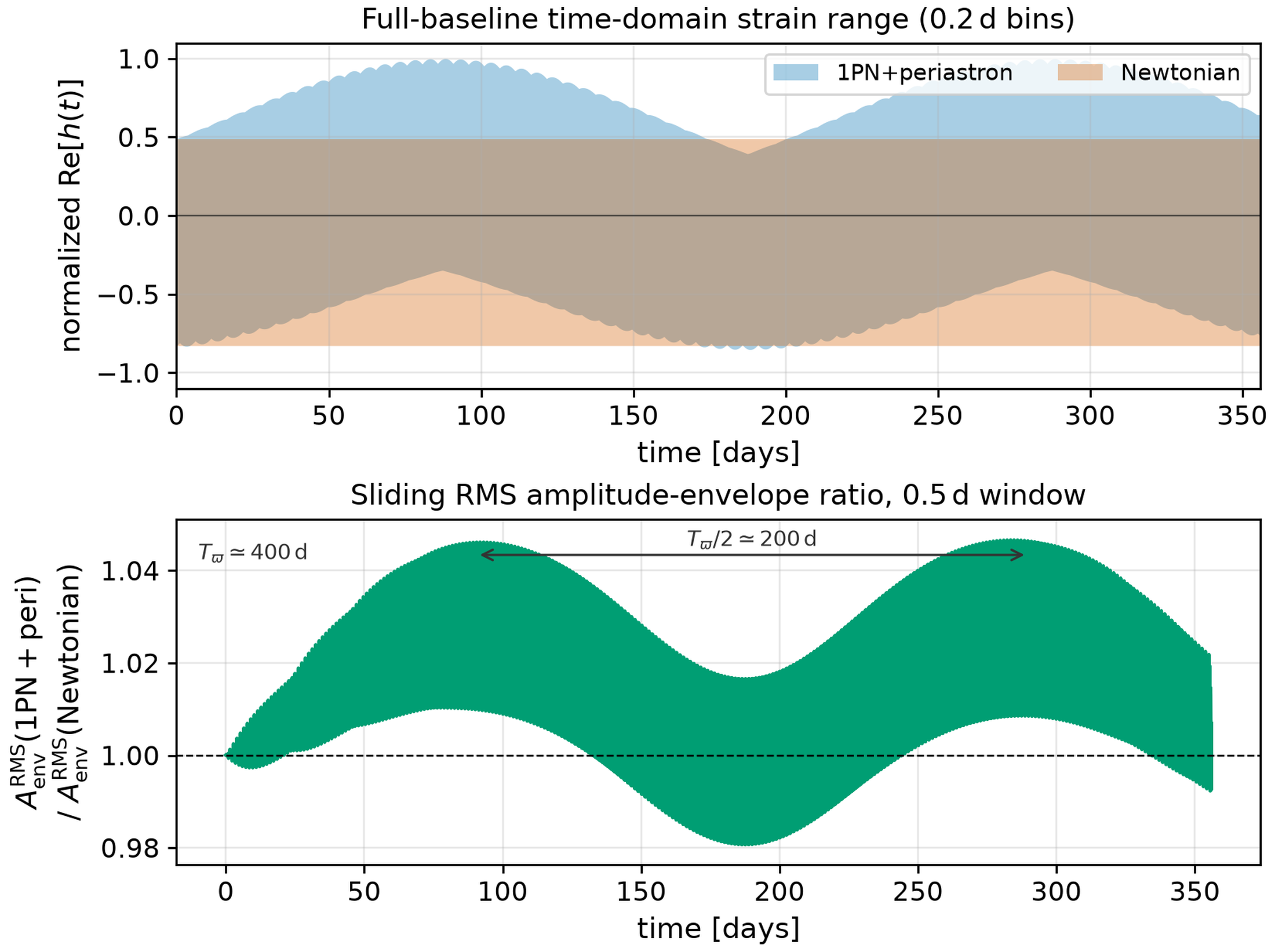}
\caption{\label{fig:periastron_envelope}
Time-domain and amplitude-envelope diagnostics for conservative 1PN
periastron advance. The source has $e=0.6$,
$m_1=m_2=1.4M_\odot$, and $f_0=1\,{\rm mHz}$, with radiation reaction
switched off. Top panel: full-baseline normalized real-strain range,
computed in $0.2\,{\rm d}$ bins, comparing the 1PN waveform with periastron
advance to the corresponding Newtonian eccentric waveform. Bottom panel:
ratio of the sliding RMS envelope of the 1PN periastron-advance waveform to
the Newtonian envelope, using a $0.5\,{\rm d}$ window. For this source the
full periastron-precession period is $T_\varpi\simeq400\,{\rm d}$; because
the quadrupolar waveform is approximately invariant under a $\pi$ rotation of
the apsidal line, the dominant envelope modulation appears on the half-period
$T_\varpi/2\simeq200\,{\rm d}$.}
\end{figure*}

In addition, the source model presented in this paper also models general relativistic periastron advance in a computationally accurate and efficient way. It turns out that 
this effect should be relevant for eccentric compact binaries that involve neutron stars and black holes. It is straightforward to compute the expected 
frequency shift of any harmonic
due to the periastron advance \cite{TessmerGopakumar2006}, and it  reads
\begin{equation}
\Delta f_k \sim
\frac{1.2\times10^{-7}}{(1-e^2)}
\left(\frac{m}{2.8M_\odot}\right)^{2/3}
\left(\frac{f}{10^{-3}\,{\rm Hz}}\right)^{5/3}
\,{\rm Hz}.
\tag{23}
\end{equation}
It is instructive to compare this frequency shift with LISA's one-year frequency resolution, $\Delta f_{\rm LISA}\sim 3\times10^{-8},{\rm Hz}$. Such a comparison indicates that the frequency shift induced by relativistic periastron advance can be resolvable by LISA, and therefore should not be neglected when modelling eccentric, massive Galactic binaries. The relevant precession timescale is set by the difference between the
azimuthal and radial frequencies. Since $\dot{\lambda}=\omega$ and
$\dot{l}=n$, the secular periastron-advance rate is
\begin{equation}
\dot{\varpi} = \omega-n = k n
= \frac{k}{1+k}\,\omega ,
\end{equation}
where $\omega=(1+k)n$. We therefore define the full periastron-precession
period as
\begin{equation}
T_\varpi = \frac{2\pi}{\dot{\varpi}}
= \frac{2\pi}{\omega-n}
= \frac{2\pi(1+k)}{k\omega}.
\label{eq:periastron_timescale}
\end{equation} 

For the parameters used in Fig.~\ref{fig:periastron_envelope}, this gives
$T_\varpi\simeq400\,{\rm d}$ while employing the 1PN-accurate expression for 
$k$, given earlier as $ k_{\rm 1PN}=\frac{3x}{1-e_t^2} $.
The dominant quadrupolar amplitude envelope is
approximately invariant under a rotation $\pi$ of the apsidal line, so the
modulation of the visible envelope occurs roughly $T_\varpi/2$.
\begin{figure*}[!tbhp]
\includegraphics[width=\textwidth]{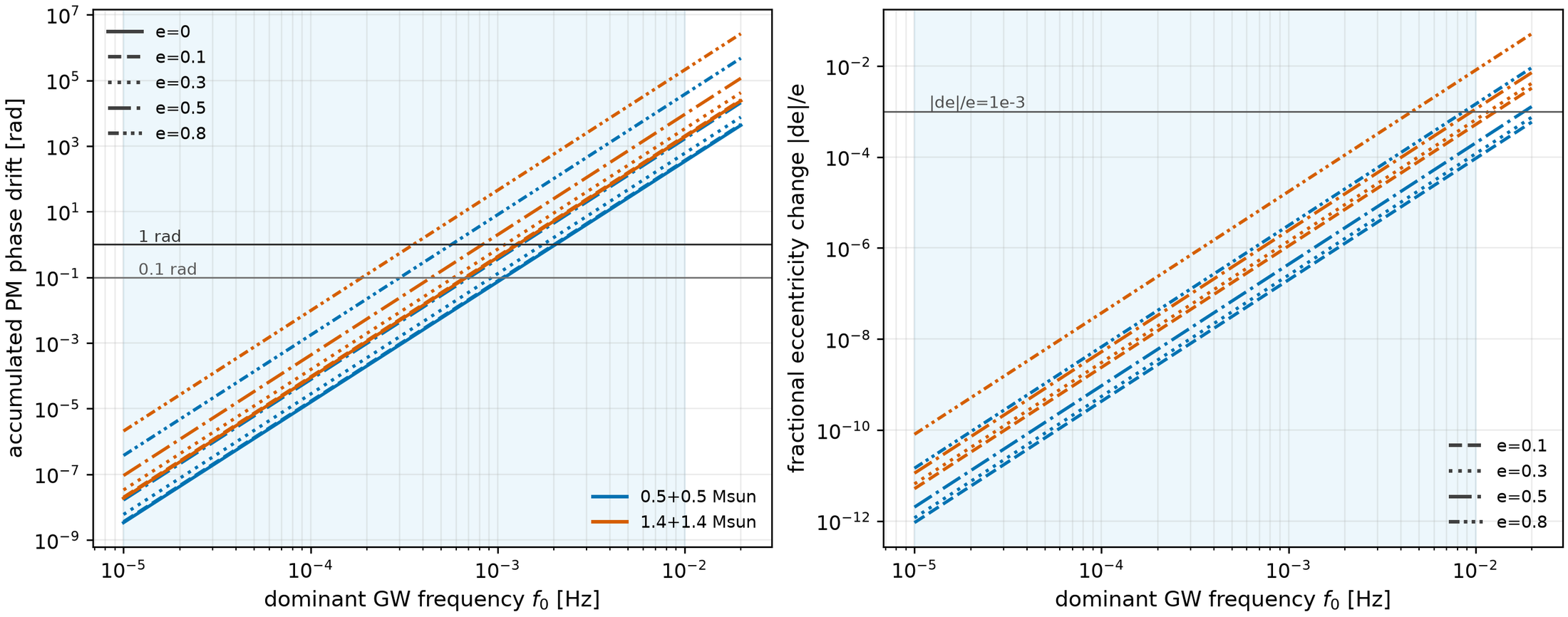}
\caption{\label{fig:pm_relevance_curves}
Peters--Mathews relevance over four years for representative binaries. The
left panel shows the accumulated orbital phase drift, which is the part of
secular evolution that can affect matched filtering because it enters through
the time integral of the orbital frequency. The right panel shows the
fractional eccentricity change, which mainly affects the relative harmonic
amplitudes and remains very small across the shaded LISA sensitivity band,
$10^{-5}$--$10^{-2}\,{\rm Hz}$. Colors indicate the two equal-mass binaries
$0.5+0.5\,M_\odot$ and $1.4+1.4\,M_\odot$; line styles indicate
eccentricities $e=0,0.1,0.3,0.5,0.8$, with the $e=0$ curve omitted from the
right panel because $|\Delta e|/e$ is not defined there.}
\end{figure*}

Along with periastron advance, gravitational radiation reaction also
produces a secular frequency drift by driving the gradual inspiral of the
binary. A close inspection of the Peters--Mathews based  equations,
Eqs.~(\ref{eq:omegadot}) and~(\ref{eq:edot}), reveals that
gravitational-wave emission induces secular temporal variations in both the
orbital angular frequency $\omega(t)$ and the eccentricity $e_t(t)$. While
both evolve over time, they affect $h_{\times,+}(t)$ in quantitatively
different ways. The time evolution of the eccentricity primarily modulates the
amplitudes of the individual harmonic modes, redistributing the GW power
across the frequency spectrum. In contrast, changes in the orbital angular
frequency $\omega(t)$ directly propagate into the orbital phase through the
integral
\begin{equation}
\lambda(t)=\int^t \omega(t')\,dt' .
\end{equation}
Since the phase of a given harmonic is a linear combination of $\lambda(t)$
and the periodic function $W(u)$, any secular drift in $\omega$ accumulates
quadratically over the observation baseline:
\begin{equation}
\lambda(t) \simeq
\lambda_0+\omega_0 t+\frac{1}{2}\dot{\omega}_0 t^2+\cdots .
\end{equation}
Consequently, even a tiny $\dot{\omega}$ can lead to a significant dephasing
after several years, substantially altering the total number of
gravitational-wave cycles accrued during that time. This sensitivity opens up
the possibility that a small but sustained $\dot{\omega}$ could produce a
detectable evolution in the gravitational-wave frequency
\cite{TessmerGopakumar2006}.

In contrast, variations in the eccentricity $e_t$ primarily modulate the
amplitudes of the individual harmonic modes of $h_{\times,+}(t)$, while also
indirectly influencing the rate of change of the orbital frequency
$\dot{\omega}$ through their coupling in the quadrupolar radiation-reaction
equations, namely Eqs.~(\ref{eq:omegadot}) and~(\ref{eq:edot}).
Furthermore, over the observational baseline relevant to Galactic
binaries, the relative change in the eccentricity is typically subdominant
compared to the accumulated phase change caused by the frequency drift. To quantify this separation of effects, we construct the differential relation
\begin{equation}
\frac{d\omega}{de_t}=\omega\,\kappa(e_t)
\end{equation}
directly from Eqs.~(\ref{eq:omegadot}) and~(\ref{eq:edot}), where
$\kappa(e_t)$ is a known function of the eccentricity. This equation is
separable and admits the standard Peters--Mathews analytic relation
\cite{PetersMathews1963,Peters1964,TessmerGopakumar2006}
\begin{widetext}
\begin{align}
\left(\frac{\omega}{\omega_0}\right)
&=
\frac{
e_0^{18/19}
\left(304+121e_0^2\right)^{1305/2299}
}
{\left(1-e_0^2\right)^{3/2}}
\times
\frac{
\left(1-e_t^2\right)^{3/2}
}
{
e_t^{18/19}
\left(304+121e_t^2\right)^{1305/2299}
}.
\end{align}
\end{widetext}
Consequently, inverting this expression numerically yields the implicit
mapping $e_t=e_t(\omega;\omega_0,e_0)$, thereby expressing the instantaneous
eccentricity as a function of the evolving orbital frequency, parameterized by
the initial conditions $\omega_0$ and $e_0$.

The corresponding one-year frequency shift from radiation reaction can be
estimated from the quadrupole orbit-averaged contribution to
$d\omega/dt$. Written in terms of the radial orbital frequency $f_r$, this
gives
\begin{widetext}
\begin{equation}
\Delta f_{\rm RR}
\sim
\frac{1.6\times10^{-9}}{(1-e^2)^{7/2}}
\left(\frac{m}{2.8M_\odot}\right)^{5/3}
\left(\frac{\eta}{0.25}\right)
\left(\frac{f_r}{10^{-3}\,{\rm Hz}}\right)^{11/3}
\left(
1+\frac{73}{24}e^2+\frac{37}{96}e^4
\right)
\,{\rm Hz}.
\label{eq:delta_f_rr}
\end{equation}
\end{widetext}
With these expressions, we can quantify the relative importance of frequency
evolution and eccentricity decay. For a
$0.5M_\odot+0.5M_\odot$ binary with $e_0=0.1$, the Peters--Mathews phase drift
is $7.7\times10^{-2}$ rad at $f_0=10^{-3}\,{\rm Hz}$, $4.35$ rad at
$3\times10^{-3}\,{\rm Hz}$, and $3.6\times10^{2}$ rad at
$10^{-2}\,{\rm Hz}$ over four years. The fractional change in eccentricity
over the same interval is only $2.0\times10^{-7}$, $3.7\times10^{-6}$, and
$9.2\times10^{-5}$, respectively. Thus the first evolution correction to
worry about is accumulated orbital phase drift, not the change in
eccentricity itself.

This conclusion is consistent with the much clearer measurements of orbital
evolution in binary-pulsar timing. For example, the Double Pulsar
PSR~J0737--3039A/B has an orbital period of $2.45\,{\rm hr}$, eccentricity
$e=0.088$, and a 16-year timing baseline, enabling the measurement of seven
post-Keplerian parameters and a test of the GR quadrupole prediction at the
$1.3\times10^{-4}$ level \cite{Kramer2021DoublePulsar}. The comparison with
LISA Galactic binaries is not contradictory. Pulsar timing tracks sharp radio
pulses over a much longer baseline, so secular orbital-period changes are
measured directly in pulse arrival times. 
In contrast, for the LISA mission
the sources are
wide, weakly evolving detached binaries observed for four years through the
detector response and instrumental noise. Since the secular phase drift scales
approximately as $T_{\rm obs}^2$, the longer timing baseline alone gives a
factor $(16/4)^2=16$ more leverage on the quadratic phase term, before
accounting for the very different measurement observable.

Figure~\ref{fig:pm_relevance_curves} summarizes this seperation over a
four-year observation. The accumulated phase drift can become appreciable
toward the upper mHz band or for larger chirp masses, while the fractional
eccentricity change remains very small for detached white-dwarf-like Galactic
binaries across most of the LISA band. The source-correction comparison below
verifies the same separation directly: evolving $e(t)$ alone produces
negligible mismatch, while the orbital phase part of Peters--Mathews
evolution becomes relevant near the high-frequency edge of the grid.

These source-level studies are implemented in the package called as \texttt{eGB-multi}. This implementation generates two
time-domain eccentric polarization states from the quasi-Keplerian orbital
variables, with the Newtonian eccentric harmonic structure as the baseline and
with conservative 1PN periastron advance and Peters--Mathews radiation
reaction available as source-level extensions. The source model is
implemented with two independent sets of switches, one on physics and the other on the evolution, the available options are:
\begin{itemize}
\item \texttt{physics\_mode="newtonian"}: Newtonian eccentric orbital motion,
with no explicit 1PN orbital corrections and no periastron advance.
\item \texttt{physics\_mode="1pn"}: the conservative 1PN source model,
including both 1PN orbital corrections and periastron advance.
\item \texttt{evolution\_mode="fixed"}: the orbital frequency and eccentricity
are held fixed at their initial values.
\item \texttt{evolution\_mode="peters\_mathews"}: both the orbital frequency
and eccentricity are evolved using the Peters--Mathews radiation-reaction
equations.
\item \texttt{evolution\_mode="peters\_mathews \newline \_orbital\_only"}: only the
orbital-frequency evolution is retained, while eccentricity is held fixed.
\item \texttt{evolution\_mode="peters\_mathews \newline \_eccentricity\_only"}: only the
eccentricity evolution is retained, while the orbital frequency is held fixed.
\end{itemize}
This modular construction allows us to isolate the dominant physical effect,
namely the multi-harmonic structure produced by eccentricity, while retaining
the subdominant secular and relativistic corrections needed for targeted
regions of parameter space.
In other words, the package allows one to choose among three levels of dynamical
description for Galactic binaries: (i) Newtonian closed
eccentric orbits, (ii) relativistically precessing eccentric
orbits incorporating the first Post Newtonian-accurate
periastron advance, and (iii) precessing and shrinking
eccentric orbits due to quadrupolar order gravitational wave 
emission, computed by Peters \& Mathews.

\section{\label{sec:lisa_response}Detector response, noise weighting, and source-correction tests}
\label{sec:statistics}

\subsection{\label{subsec:egbmulti_lisa}LISA response implementation in \texttt{eGB-multi}}

The source-level model described in Sec.~\ref{sec:model} produces the two
gravitational-wave polarizations $h_+(t)$ and $h_\times(t)$ at the Solar
System barycenter.  To compare such a source with LISA data, these
polarizations must be projected onto the moving detector constellation.  In
\texttt{eGB-multi}, this is done directly in the time domain by first
constructing the six one-way inter-spacecraft Doppler measurements and then
combining them into time-delay interferometry (TDI) observables.

For each directed link from spacecraft $j$ to spacecraft $i$, we denote the
unit arm direction by $\hat{n}_{ij}$ and the light-travel time by $L_{ij}$.
The gravitational wave is evaluated at the retarded emission and reception
times, $t_{\rm send}$ and $t_{\rm rec}$, along that link.  The schematic
one-way Doppler response is
\begin{equation}
y_{ij}(t)
=
\frac{1}{2}
\frac{
\hat{n}_{ij}^a \hat{n}_{ij}^b
\left[
h_{ab}(t_{\rm send})-h_{ab}(t_{\rm rec})
\right]
}{
1-\hat{n}_{ij}\cdot\hat{k}
},
\label{eq:one_way}
\end{equation}
where $\hat{k}$ is the gravitational-wave propagation direction.  The
implementation follows the LISA science ground-segment conventions document
\cite{LISAConventions2026}, in which the label $ij$ denotes a measurement
received at spacecraft $i$ and emitted by spacecraft $j$.

The six one-way links are then combined into second-generation Michelson TDI
observables $X$, $Y$, and $Z$ using \texttt{pyTDI}.  We finally rotate these
Michelson observables to the noise-orthogonal $A,E,T$ basis,
\begin{equation}
\begin{aligned}
A&=\frac{Z-X}{\sqrt{2}},
&
E&=\frac{X-2Y+Z}{\sqrt{6}},
\\
T&=\frac{X+Y+Z}{\sqrt{3}}.
\end{aligned}
\end{equation}
The $T$ channel is generated and stored as a consistency check, but the
fitting factors and residuals quoted below use only the $A$ and $E$ channels.

The response-level implementation in \texttt{eGB-multi} is organized as a
pipeline:
\begin{itemize}
\item The source polarizations $h_+(t)$ and $h_\times(t)$ are combined with
the sky location and polarization basis to construct the transverse-traceless
metric perturbation $h_{ab}(t)$.
\item The LISA constellation is evaluated on the requested time grid using the
analytic trailing-orbit model supplied by \texttt{lisaorbits}.
\item The six exact retarded one-way Doppler links $y_{ij}(t)$ are computed in
the time domain.
\item The link measurements are passed to \texttt{pyTDI} to form
second-generation Michelson $X,Y,Z$ observables.
\item The $X,Y,Z$ channels are converted to $A,E,T$, and the $A,E$ channels
are used for the noise-weighted inner products.
\item The same response path can be applied to any source branch:
\texttt{newtonian}, \texttt{1pn}, \texttt{fixed}, or
\texttt{peters\_mathews}.
\end{itemize}

For the response-level results in this paper we use this exact retarded-link
time-domain path rather than a reduced response approximation.  The LISA
geometry is precomputed on the common time grid and reused across source
batches, while the source and link evaluations are organized in fixed-size
arrays for JAX acceleration where possible.  This allows us to change the
source physics without changing the detector response, TDI convention, or
noise-weighting prescription.

\subsection{\label{subsec:noise_inner_products}Noise weighting and waveform comparisons} 
The weighting curve used in the current numerical experiments is estimated
from simulated LISA instrumental noise generated with the \texttt{lisainstrument}
package and projected through \texttt{pyTDI} \cite{PyTDI}. The resulting $A,E,T$ noise
channels are converted to a Welch power spectral density estimate. The same
formalism can be extended to include analytic instrumental-noise and confusion
foreground models \cite{RobsonCornishLiu2019,Karnesis2021}.
The enabled \texttt{lisainstrument} noise terms are the package defaults:
laser frequency noise, clock and modulation noise, test-mass acceleration
noise, optical-metrology noise in the science, test-mass, and reference
interferometers, backlink and reciprocal-backlink noise, ranging noise,
MOSA longitudinal jitter, spacecraft and MOSA angular jitter, differential
wavefront-sensing readout noise, tilt-to-length coupling, and MOC
time-correlation noise. No Galactic confusion foreground, stochastic
astrophysical background, glitches, or injected gravitational-wave foreground
is included in this PSD estimate.

\begin{figure}[!htbp]
\includegraphics[width=\columnwidth]{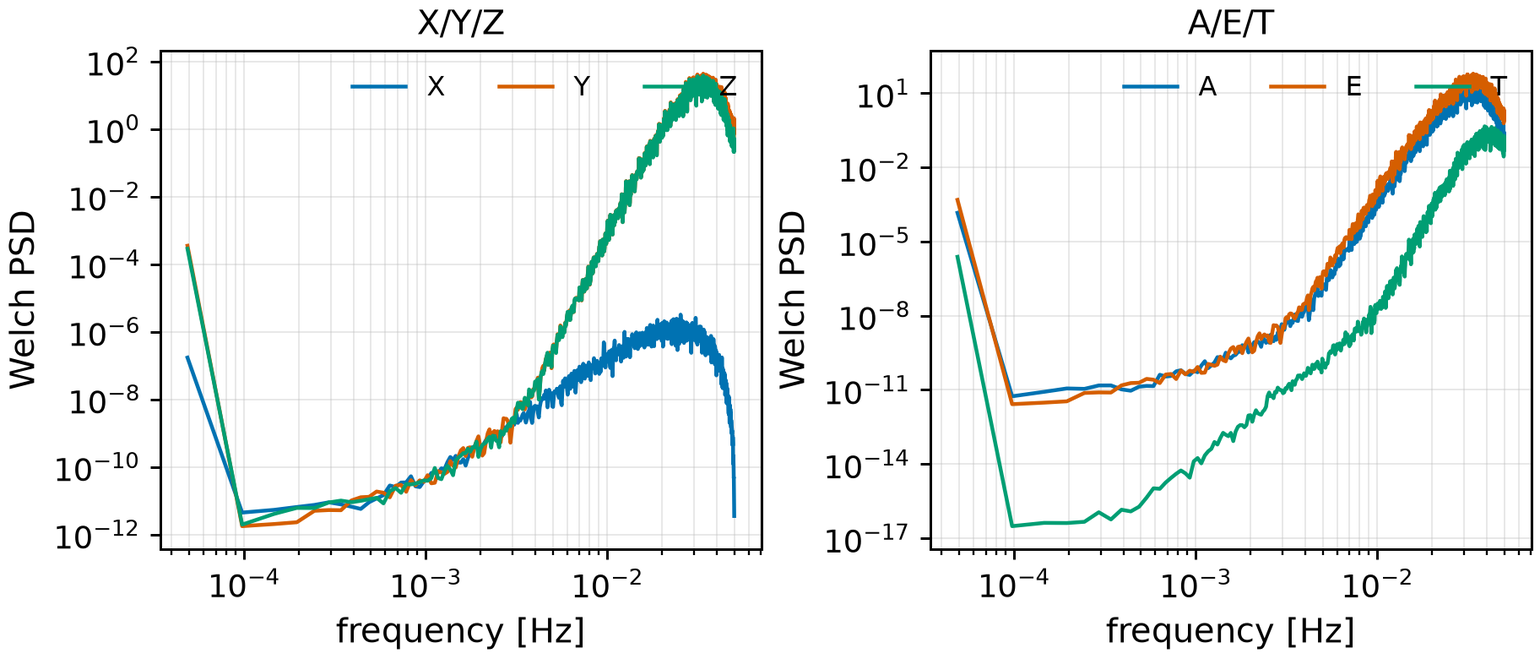}
\caption{\label{fig:welch_psd}
Welch PSD estimated from a \texttt{lisainstrument} simulation and used as the
noise-weighting curve in the response-level fitting-factor calculations. The noise
simulation used 8192 samples at $\Delta t=10\,{\rm s}$ and Welch segments of
2048 samples, giving frequency spacing $4.8828125\times10^{-5}\,{\rm Hz}$.
The inner products use the $A$ and $E$ PSDs only.}
\end{figure}

\begin{figure*}[!tbhp]
\includegraphics[width=\textwidth]{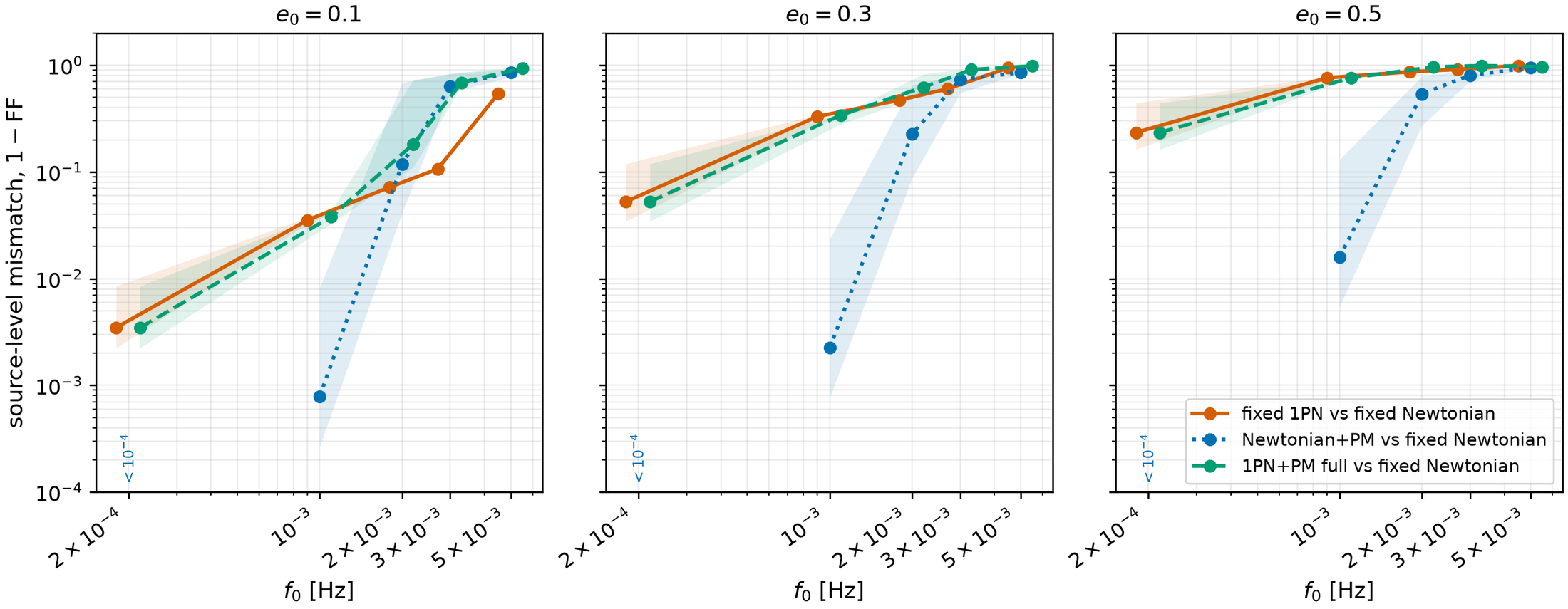}
\caption{\label{fig:subdominant_corrections}
Noise-weighted source-level mismatches from optional eccentric-binary source
corrections over four years.  The baseline eccentric model in this paper is
the fixed 1PN quasi-Keplerian waveform, with conservative periastron advance
included.  The Newtonian comparison shows the size of omitting these
conservative 1PN effects, while the Peters--Mathews comparison shows the
additional mismatch obtained by evolving the 1PN source under radiation
reaction.  Points below $10^{-4}$ are labelled rather than shown on the
logarithmic scale.  The dominant radiation-reaction contribution comes from
orbital-frequency evolution; eccentricity decay is negligible over the
Galactic-binary baselines considered here.}
\end{figure*}

We quantify the loss incurred by circular templates using noise-weighted
fitting factors and relative residuals. For Fourier-domain waveforms
$\tilde{a}(f)$ and $\tilde{b}(f)$, the one-sided inner product is
\begin{equation}
(a|b)=
4\,{\rm Re}\int_{0}^{\infty}
\frac{\tilde{a}(f)\tilde{b}^*(f)}{S_n(f)}\,df .
\label{eq:inner_product}
\end{equation}
For response-level comparisons, the inner product is summed over the $A$ and
$E$ TDI channels,
\begin{equation}
(a|b)_{AE}=(a_A|b_A)+(a_E|b_E).
\end{equation}
Thus each source configuration produces one combined $A/E$ fitting factor;
the medians reported below are taken over source configurations, not over
TDI channels. The fixed-grid fitting factor is
\begin{equation}
{\rm FF}
=
\frac{|(d_{\rm ecc}|h_{\rm circ})|}
{\sqrt{(d_{\rm ecc}|d_{\rm ecc})(h_{\rm circ}|h_{\rm circ})}},
\label{eq:ff}
\end{equation}
where the numerator is evaluated after summing the $A$ and $E$ contributions.
Here $d_{\rm ecc}$ denotes the eccentric signal and $h_{\rm circ}$ the circular
template.
The mismatch is $1-{\rm FF}$. 

\subsection{\label{subsec:subdominant}Relevance of optional source corrections}

We now use the PSD-weighted inner product defined above to quantify the
importance of the optional source corrections in \texttt{eGB-multi}.  In the
paper, our baseline eccentric source model is the fixed 1PN quasi-Keplerian
waveform, where ``1PN'' always includes conservative periastron advance.  The
remaining optional source correction is then the secular Peters--Mathews
radiation-reaction evolution of the orbital frequency and eccentricity.  This
choice separates two questions: first, how different is a reduced Newtonian
eccentric source from the 1PN precessing eccentric baseline; and second, when
does Peters--Mathews evolution need to be added on top of the fixed 1PN model?

Figure~\ref{fig:subdominant_corrections} shows this hierarchy at the source
level over four-year baselines.  We compare reduced models against more
complete eccentric waveforms for representative binaries spanning frequency,
eccentricity, and mass.  The fixed Newtonian comparison measures the size of
the conservative 1PN correction, including the periastron-induced difference
between the radial and azimuthal orbital phases.  The Peters--Mathews
comparison instead keeps the 1PN precessing dynamics fixed and asks how much
mismatch is produced by allowing the binary to evolve secularly under
quadrupolar radiation reaction.

The dominant Peters--Mathews contribution is the orbital-frequency evolution.
Its effect accumulates in the phase through
$\lambda(t)=\int \omega(t)\,dt$, so even a small $\dot{\omega}$ can become
important over multi-year observations.  By contrast, the direct decay of the
eccentricity is subdominant for the Galactic-binary systems considered here:
over LISA baselines it changes the harmonic amplitudes only weakly compared to
the accumulated phase drift from orbital evolution.  Therefore the correction
that must be monitored in practice is not usually $e_t(t)$ itself, but the
mismatch between the fixed 1PN waveform and the 1PN waveform with full
Peters--Mathews evolution.

This source-correction study is also the basis for the automatic evolution
mode implemented in \texttt{eGB-multi}.  The user specifies a mismatch
tolerance $\epsilon_{\rm PM}$ relative to the 1PN waveform with full
Peters--Mathews evolution.  For each source and observation time, the package
evaluates or predicts
\begin{equation}
1-{\rm FF}
\left[
h_{\rm 1PN+PM},
h_{\rm 1PN,fixed}
\right].
\end{equation}
If this quantity exceeds the requested tolerance, Peters--Mathews evolution is
enabled; otherwise the fixed 1PN waveform is used.  In the package this is the
\texttt{evolution\_mode="auto"} option.  The rule can be evaluated directly
for a single source, or calibrated once on a representative grid and reused in
large population studies.  This makes the source model tolerance-aware: fast
exploratory runs can use a looser threshold, while high-accuracy studies can
lower $\epsilon_{\rm PM}$ and automatically retain radiation reaction wherever
it is needed.

\begin{figure}[!htbp]
\includegraphics[width=\columnwidth]{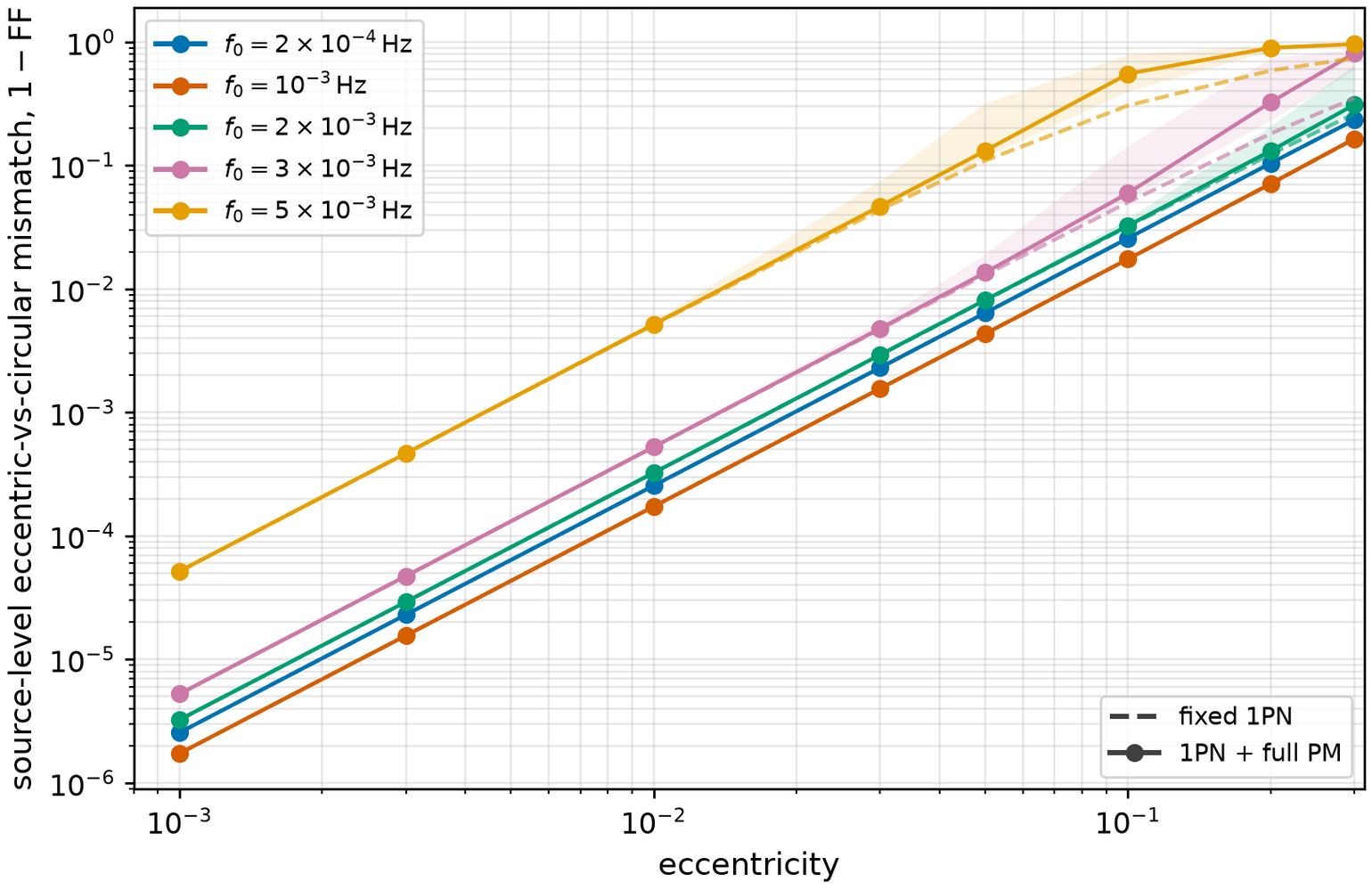}
\caption{\label{fig:full_model_vs_fixed}
Source-level eccentric-vs-circular mismatch for the fixed 1PN eccentric
baseline and for the same 1PN source with full Peters--Mathews evolution.
In this paper, 1PN always includes conservative periastron advance.  The
comparison shows that the dominant eccentric-vs-circular mismatch is produced
by eccentric harmonic content, while Peters--Mathews evolution mainly affects
the high-frequency region where secular orbital-phase drift accumulates over
the observation time.}
\end{figure}

Figure~\ref{fig:full_model_vs_fixed} gives the complementary
eccentric-vs-circular comparison.  Here the eccentric source is compared
against the circular approximation, so the plotted mismatch includes the
dominant loss caused by missing eccentric harmonic structure.  The fixed 1PN
eccentric waveform already captures this leading effect.  Adding full
Peters--Mathews evolution modifies mainly the high-frequency part of the grid,
where the secular phase drift becomes appreciable.  Thus the response-level
study below uses the fixed 1PN eccentric waveform as the controlled baseline,
while the automatic Peters--Mathews switch provides the refinement required by
a chosen mismatch tolerance.

\section{\label{sec:results}Circular and eccentric comparison}
In this section we quantify the impact of orbital eccentricity on
LISA searches for Galactic binaries when quasi-circular templates are
used. We first compare eccentric signals against circular template
families in order to identify the region of parameter space where
eccentricity produces a measurable loss in match. We then compare our
time-domain eccentric response with existing circular Galactic-binary
response models. This second comparison serves two purposes: it provides
a circular-limit consistency check for the implementation, and it places
the computational cost of the eccentric calculation in context relative
to established fast circular-response models.

To carry out this comparison, we construct a response-level mismatch
grid in which the eccentric signal is generated with the full
time-domain LISA response, while the template family is restricted to
quasi-circular binaries. Our wide response-level grid spans initial orbital frequency, component
masses, sky position, inclination, and eccentricity, and uses the
automatic source-evolution prescription introduced above.

The grid uses a one-year observation,
$T_{\rm obs}=365.25\,{\rm d}$, sampled at $\Delta t=50\,{\rm s}$, giving
631152 time samples. The grid spans
\begin{equation}
\begin{aligned}
f_0=\{&10^{-4},2\times10^{-4},5\times10^{-4},
10^{-3},\\
&2\times10^{-3},3\times10^{-3}\}\,{\rm Hz},
\end{aligned}
\end{equation}
and eccentricities
\begin{equation}
\begin{aligned}
e_0=\{&0,10^{-3},3\times10^{-3},10^{-2},
3\times10^{-2},5\times10^{-2},\\
&0.1,0.2,0.3,0.5\}.
\end{aligned}
\end{equation}
For each point we also vary three inclinations
$\iota=\{0.2,0.8,1.2\}$, four ecliptic sky locations, and three mass pairs
$(0.5,0.5)$, $(0.8,0.6)$, and $(1.0,0.3)$ in solar masses, giving 2160
signal-template comparisons. Each eccentric signal is propagated through the
exact retarded-link response, converted to second-generation TDI, and compared
in the $A,E$ inner product defined in Sec.~\ref{sec:lisa_response}.

The eccentric signal is generated with the baseline 1PN source model, including
periastron advance, and with the automatic Peters--Mathews evolution switch
enabled at mismatch tolerance $10^{-2}$. No $\dot f$ is imposed on
the eccentric signal. Instead, the package decides whether the fixed-orbit
source is sufficient or whether the full Peters--Mathews evolution should be
used. In this run the auto mode selects the fixed source for 2004 cases and
Peters--Mathews evolution for 156 cases. The switch is inactive below
$2\times10^{-3}\,{\rm Hz}$, turns on for the $e_0=0.5$ cases at
$2\times10^{-3}\,{\rm Hz}$, and becomes more common at
$3\times10^{-3}\,{\rm Hz}$.

The circular template family is deliberately generous. For each circular
template we set $e_0=0$ and maximize over a grid of circular frequency
derivatives,
\begin{equation}
\dot f_{\rm circ} =
\{0,0.25,0.5,1,2,4,8,16,32\}\dot f_{\rm GR},
\end{equation}
where $\dot f_{\rm GR}$ is the vacuum-GR circular chirp implied by the
component masses and $f_0$. Thus any remaining mismatch is not simply the
failure to include a circular chirp; it is the part of the eccentric signal
that cannot be absorbed by the usual circular template structure.

The central response-level result is shown in
Fig.~\ref{fig:tdi_mismatch}. Each curve gives the median mismatch over sky
position, inclination, and mass at fixed $f_0$ and $e_0$. The $e_0=0$ points are not shown:
they provide a useful circular-limit check, but their mismatch is at numerical
precision.

\begin{figure}[!htbp]
\includegraphics[width=\columnwidth]{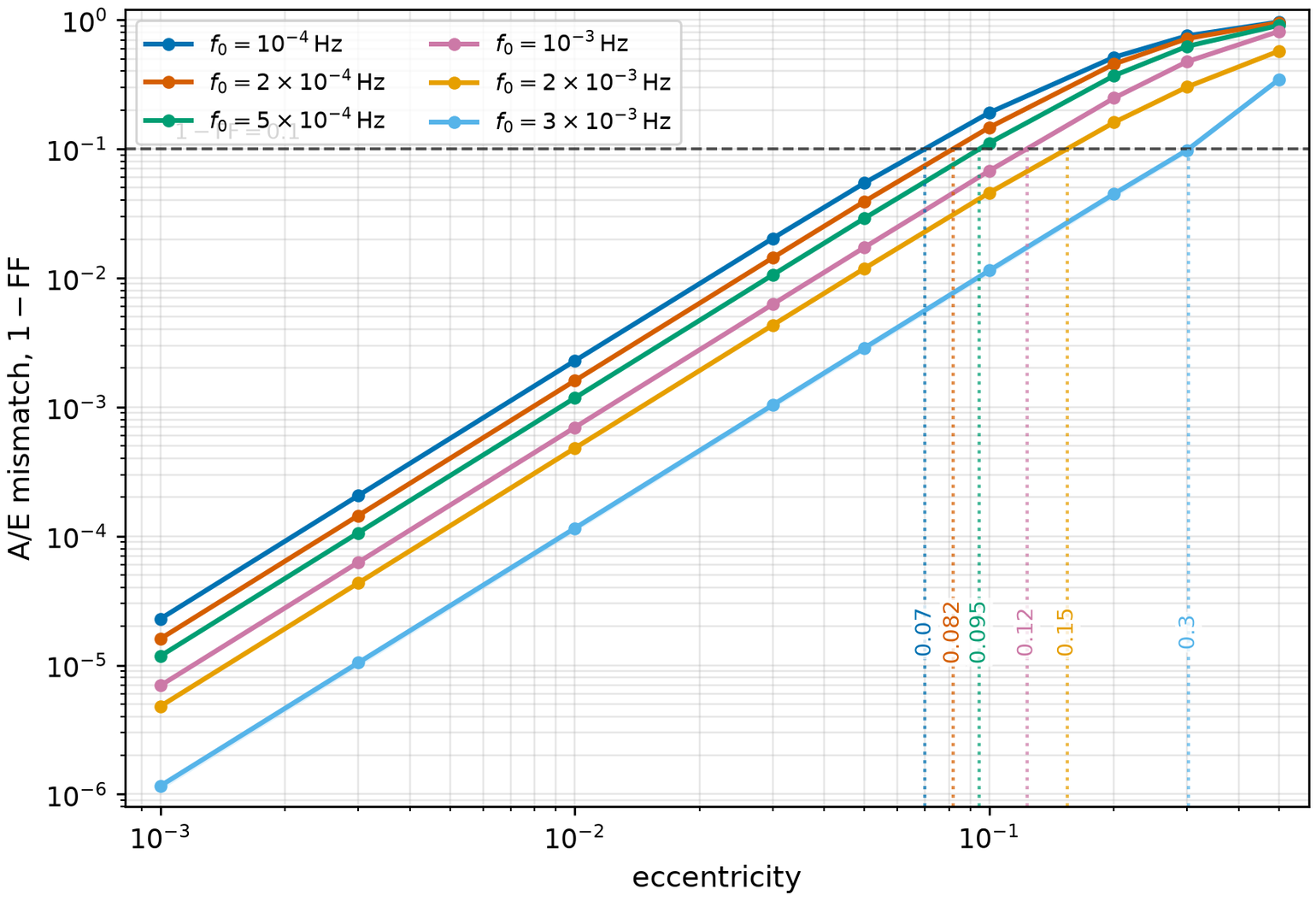}
\caption{\label{fig:tdi_mismatch}
Response-level eccentric-vs-circular mismatch after exact time-domain LISA
links, second-generation TDI, and projection to the $A,E$ channels. The
eccentric signal uses the baseline 1PN model with periastron advance and the
automatic Peters--Mathews evolution switch at tolerance $10^{-2}$; the circular
template has $e_0=0$ and is maximized over
$\dot f_{\rm circ}/\dot f_{\rm GR}=\{0,0.25,0.5,1,2,4,8,16,32\}$.
Points show the median over sky position, inclination, and mass.
The horizontal dashed line marks $1-{\rm FF}=0.1$, and the vertical dotted
lines mark the eccentricity at which each frequency track reaches that
threshold.}
\end{figure}

The median mismatch grows steeply with eccentricity. At fixed eccentricity the mismatch also depends on frequency through the LISA
response, Doppler modulation, TDI transfer function, and noise weighting. For
example, at $e_0=0.1$ the median mismatch decreases from
$1.91\times10^{-1}$ at $10^{-4}\,{\rm Hz}$ to $1.14\times10^{-2}$ at
$3\times10^{-3}\,{\rm Hz}$. Table~\ref{tab:eccentricity_threshold_0p1} summarizes the eccentricity at
which the median response-level mismatch with quasi-circular templates reaches
\(1-\mathrm{FF}=0.1\), shown separately for the sampled component-mass pairs.

\begin{table}[t]
\centering
\caption{
Eccentricity threshold at which the median response-level mismatch between
eccentric signals and quasi-circular templates reaches
\(1-\mathrm{FF}=0.1\). For each initial frequency and mass pair, the median is
taken over the sampled sky position and inclination points. The threshold
eccentricity is obtained by log--log interpolation between adjacent sampled
eccentricities, matching the vertical markers in Fig.~\ref{fig:tdi_mismatch}.
}
\label{tab:eccentricity_threshold_0p1}
\begin{ruledtabular}
\begin{tabular}{cccc}
\(f_0\,[{\rm Hz}]\) &
\multicolumn{3}{c}{\(e_0\) at \(1-\mathrm{FF}=0.1\)} \\
\cline{2-4}
&
\(0.5+0.5\,M_\odot\) &
\(0.8+0.6\,M_\odot\) &
\(1.0+0.3\,M_\odot\) \\
\hline
\(1.0\times10^{-4}\) & \(0.0699\) & \(0.0699\) & \(0.0699\) \\
\(2.0\times10^{-4}\) & \(0.0818\) & \(0.0818\) & \(0.0818\) \\
\(5.0\times10^{-4}\) & \(0.0947\) & \(0.0947\) & \(0.0947\) \\
\(1.0\times10^{-3}\) & \(0.1231\) & \(0.1232\) & \(0.1232\) \\
\(2.0\times10^{-3}\) & \(0.1540\) & \(0.1540\) & \(0.1540\) \\
\(3.0\times10^{-3}\) & \(0.3028\) & \(0.3024\) & \(0.3028\) \\
\end{tabular}
\end{ruledtabular}
\end{table}

The response-level calculation above deliberately uses the exact time-domain
retarded-link implementation in \texttt{eGB-multi}, rather than a reduced
fast-response approximation.  For each source, the package evaluates the
eccentric polarizations on the required time grid, projects them onto the six
one-way LISA links, and then passes these link observables to \texttt{pyTDI}
to construct second-generation Michelson variables and the $A,E$ channels.
The LISA orbit and link geometry are precomputed on the observation grid, while
the fixed-orbit eccentric response is evaluated with JAX on fixed array shapes.
When the automatic Peters--Mathews switch selects an evolved source, the same
retarded-link and TDI pathway is used, with the source evolution evaluated at
the relevant retarded times.  Thus the mismatch curves in
Fig.~\ref{fig:tdi_mismatch} do not rely on a monochromatic, rigid-adiabatic, or
frequency-domain approximation to the detector response.

This exact time-domain implementation is intentionally conservative for the
present study.  It is slower than specialized circular Galactic-binary
responses, but it gives a direct reference calculation for eccentric binaries
with unequal-arm delays, time-dependent LISA geometry, and time-domain TDI. A useful comparison point is \texttt{JAXGB}, which belongs to the same fast
Galactic-binary response lineage as \texttt{fastGB} and \texttt{GBGPU}
\cite{FastGB,JAXGB}.  Its more accurate windowed response mode is closely
related to the time-domain-calibrated response construction of
Ref.~\cite{Riegger2024}.

We then compare the \texttt{eGB-multi} time-domain response with the fast
circular response implemented in \texttt{JAXGBaccurate}. This comparison is
used mainly to place the eccentric calculation in context. In convention-controlled circular tests, where the source phase, polarization
conventions, transfer response, and TDI construction are made identical, the
\texttt{eGB-multi} circular response agrees with the corresponding fast
circular response at numerical precision. The raw comparison with
\texttt{JAXGBaccurate}, however, retains a nonzero circular-limit baseline.
We interpret this baseline as a response-model difference associated with
comparing the brute-force time-domain pyTDI construction used in
\texttt{eGB-multi} to the native sparse/windowed fast-response construction
used by \texttt{JAXGBaccurate}.

Figure~\ref{fig:jaxgbaccurate_comparison} keeps this response-model difference
explicit. The \(e_0=0\) points show the circular-limit baseline between the two
response models, while the growth with eccentricity shows the additional
mismatch introduced by eccentric source dynamics. The wall-time comparison in
the right panel illustrates the cost of the brute-force time-domain response:
the dominant additional expense in \texttt{eGB-multi} is the construction of
the retarded one-way link waveforms before forming the TDI observables. For the single-source response benchmark shown here, the brute-force
\texttt{eGB-multi} time-domain response is typically a factor of
\(\sim 2\)--\(4\) slower than \texttt{JAXGBaccurate}, with a median slowdown of
about \(2.7\) over the sampled frequencies and eccentricities. 
\begin{figure*}[t]
\centering
\includegraphics[width=\textwidth]{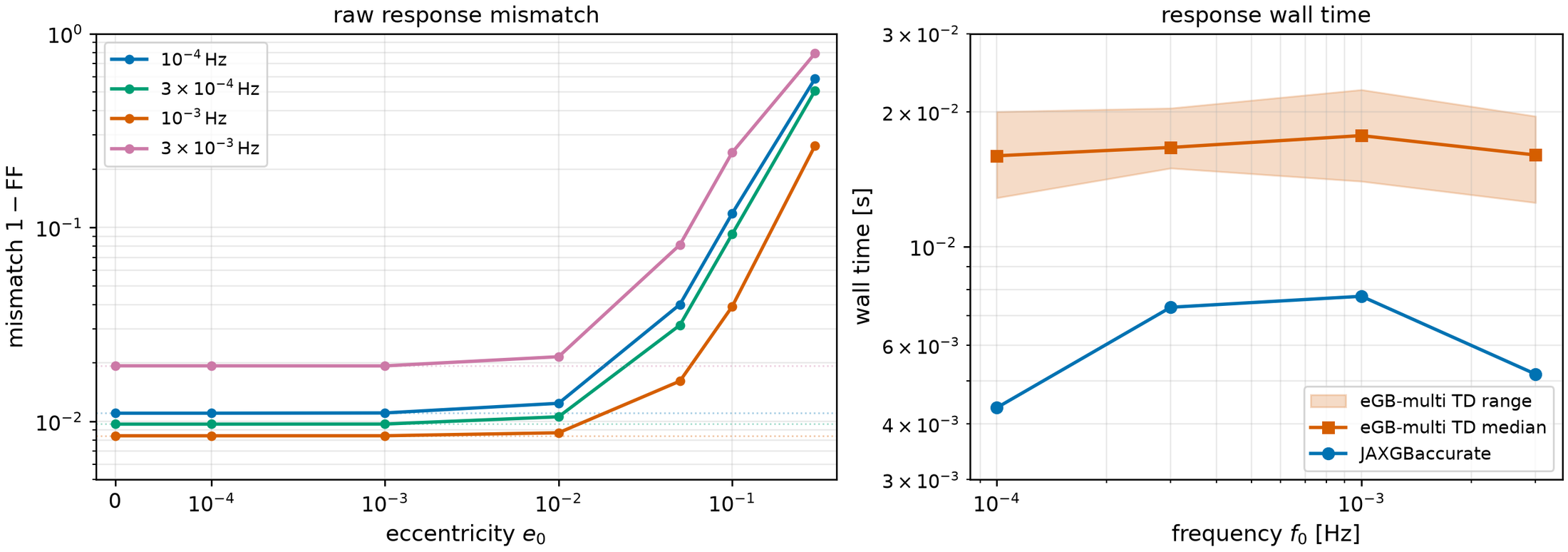}
\caption{
Comparison between the \texttt{eGB-multi} time-domain response and the circular
\texttt{JAXGBaccurate} response for a
\(0.5M_\odot+0.5M_\odot\) binary. The left panel shows the raw response-level
mismatch \(1-\mathrm{FF}\), retaining the nonzero circular-limit baseline at
\(e_0=0\). We interpret this baseline as a response-model difference between
the brute-force time-domain pyTDI construction and the native sparse/windowed
\texttt{JAXGBaccurate} response. Dotted horizontal lines mark this baseline for
each frequency. The right panel compares the wall time of the fast circular
response with the brute-force \texttt{eGB-multi} time-domain response. The
shaded band gives the range of \texttt{eGB-multi} timings over the sampled
eccentricities, and the solid orange curve gives the median. All wall-clock timings reported in this work were obtained on an Apple M4 arm64 machine with 16 GB memory running macOS 26.5.2.
}
\label{fig:jaxgbaccurate_comparison}
\end{figure*}
If one wants to build a frequency-domain eccentric construction based on
circular fast responses, this would require an evaluation and combination of a
set of eccentric harmonics, each with consistent finite-arm transfer functions,
Doppler phases, polarization conventions, and TDI combinations. The number of
required harmonics grows with eccentricity. In order to retain \(99.9\%\) of
the source power, we require harmonics up to \(k=4\) at \(e_0=0.1\) and up to
\(k=6\) at \(e_0=0.3\). This can reduce the apparent speed advantage of a
harmonic-by-harmonic frequency-domain eccentric implementation before
accounting for the additional bookkeeping required by realistic LISA response
and TDI conventions. The complication is further increased once periastron
advance or secular radiation-reaction evolution is included, since the harmonic
phases are no longer described by a single fixed orbital frequency. In that
case, the frequency-domain construction must consistently track the evolving
orbital phase, periastron phase, eccentricity, and harmonic frequencies, making
a direct fast-response generalization less straightforward than in the
quasi-circular monochromatic limit.

\section{\label{sec:conclusion}Discussion and Conclusions}
We have constructed and implemented a time-domain eccentric Galactic-binary
waveform model for LISA in the \texttt{eGB-multi} package. The implementation
combines eccentric source dynamics with the time-domain LISA response by
evaluating the signal on the retarded one-way links and then forming the TDI
observables. This provides a direct response-level framework for studying
eccentric Galactic binaries without reducing the detector response to a
monochromatic or frequency-domain approximation.

A central feature of \texttt{eGB-multi} is that the level of source accuracy can
be selected according to the needs of the analysis. The package can generate 
accurate and efficient time domain 
quadruopolar order gravitational wave polarization states 
associated with Galactic binaries. Further, 
the user can choose among three levels of dynamical
description for Galactic binaries. This includes (i) Newtonian closed
eccentric orbits, (ii) relativistically precessing eccentric
orbits incorporating the first Post Newtonian-accurate
periastron advance, and (iii) precessing and shrinking
eccentric orbits due to 
Peters--Mathews radiation-reaction evolution.
Additionally, the user may invoke 
an automatic
evolution mode in which the Peters--Mathews corrections are activated only when a
user-specified mismatch tolerance is exceeded. This makes the implementation
useful both as a conservative reference calculation and as a flexible tool for
larger parameter-space studies where unnecessary source corrections should not
be included everywhere.

At the response level, we find that quasi-circular templates begin to lose
significant match once the eccentricity is sufficiently large, with the
threshold depending mainly on the initial orbital frequency. This result
remains after including the time-domain LISA response, second-generation TDI,
realistic instrumental-noise weighting, and a circular-template maximization
over the frequency derivative. The automatic evolution mode follows directly
from the same source-correction study: it keeps the cheaper fixed-source model
where radiation reaction is negligible and activates Peters--Mathews evolution
only in the part of parameter space where the requested mismatch tolerance
requires it.

We also compared the \texttt{eGB-multi} response with existing fast circular
Galactic-binary response models. In convention-controlled circular tests, where
the source phase, polarization conventions, transfer response, and TDI
construction are made identical, the circular response agrees with the
corresponding fast circular response at numerical precision. The raw comparison
with \texttt{JAXGBaccurate} retains a small response-model difference associated
with comparing the brute-force time-domain pyTDI construction used here to the
native sparse/windowed fast-response construction of \texttt{JAXGBaccurate}.
For the single-source benchmark considered in this paper, the time-domain
\texttt{eGB-multi} response is typically a factor of \(\sim2\)--\(4\) slower
than \texttt{JAXGBaccurate}. This additional cost is expected: the eccentric
calculation explicitly constructs the retarded one-way link waveforms before
forming the TDI observables.

\texttt{eGB-multi}  provides the waveform side of an
eccentric Galactic-binary analysis: PN eccentric source polarizations, optional
Peters--Mathews evolution, automatic source-correction switching, exact
time-domain one-way LISA links, \texttt{pyTDI} \(X,Y,Z\), \(A,E,T\) channels,
and PSD-weighted overlaps and mismatches.  A natural next step is to promote
these ingredients into a sampler integrated  likelihood interface, in which fixed
data streams, PSDs, frequency masks, cached LISA geometry, and TDI settings are
initialized once, while the source parameters are varied repeatedly through
\(\theta\rightarrow h_{A,E,T}(\theta)\rightarrow d-h(\theta)\rightarrow
\ln\mathcal{L}\). Doing this will also make it possible to quantify how precisely
LISA can measure eccentricity and other parameters like total mass for Galactic binaries. We leave this parameter-estimation study, including the
resulting eccentricity, total mass  measurement precision, to a future work.

This formulation is also extensible. The same architecture
can be scaled to include additional source physics, such as higher-order PN
terms, stronger secular evolution, spin effects, or environmental corrections,
while keeping the detector response and TDI construction fixed. It is also not
limited to Galactic binaries. Stellar-origin black-hole binaries observed by
LISA can have stronger evolution, higher harmonics, eccentricity, and possibly
precession, making a direct time-domain response strategy similarly useful for
building controlled reference waveforms and testing faster approximations.

The next step developments can be done in two ways. First, the response-level eccentric
template study should be extended to realistic Galactic-binary populations and
eventually to search and parameter-estimation pipelines. Second, the cost of
the exact retarded-link calculation should be reduced. A promising route is to
adapt fast time-domain factorization ideas, such as the tri-linear
representation of Ref.~\cite{Andersson2022}, and related time-domain-calibrated
response constructions \cite{Riegger2024}, as controlled approximations that
separate the slowly varying LISA geometry from the source harmonics while
remaining compatible with time-domain TDI. These developments would allow the
flexibility of the present eccentric response model to be retained while making
it practical for large-scale LISA data-analysis applications.
\begin{acknowledgments}
 S.T. and P.J. are partially supported by the Tomalla Foundation. A. G. acknowledges the support of the Department of Atomic Energy,
Government of India, under
Project ID No. RTI 4002, and 
 the CAS President’s International Fellowship Initiative (2026PVA0020). The numerical work in this work used
\texttt{lisainstrument}, \texttt{lisaorbits}, \texttt{lisa\_gw\_response},
\texttt{pyTDI}, \texttt{JAXGB}, \texttt{fastGB}, NumPy, SciPy, JAX, and Matplotlib. The authors used OpenAI Codex as an AI-assisted programming to  generate and debug plotting and benchmarking scripts, all of which is then vetted and confirmed by the authors.

\end{acknowledgments}

\bibliographystyle{apsrev4-2}
\bibliography{apssamp}

@misc{AmaroSeoane2017,
    author = "Amaro-Seoane, Pau and others",
    collaboration = "LISA",
    title = "{Laser Interferometer Space Antenna}",
    eprint = "1702.00786",
    archivePrefix = "arXiv",
    primaryClass = "astro-ph.IM",
    month = "2",
    year = "2017"
}

@article{PetersMathews1963,
    author = "Peters, P. C. and Mathews, J.",
    title = "{Gravitational radiation from point masses in a Keplerian orbit}",
    doi = "10.1103/PhysRev.131.435",
    journal = "Phys. Rev.",
    volume = "131",
    pages = "435--439",
    year = "1963"
}

@article{Peters1964,
    author = "Peters, P. C.",
    title = "{Gravitational Radiation and the Motion of Two Point Masses}",
    doi = "10.1103/PhysRev.136.B1224",
    journal = "Phys. Rev.",
    volume = "136",
    pages = "B1224--B1232",
    year = "1964"
}

@article{Kramer2021DoublePulsar,
    author = "Kramer, M. and others",
    title = "{Strong-Field Gravity Tests with the Double Pulsar}",
    eprint = "2112.06795",
    archivePrefix = "arXiv",
    primaryClass = "astro-ph.HE",
    doi = "10.1103/PhysRevX.11.041050",
    journal = "Phys. Rev. X",
    volume = "11",
    number = "4",
    pages = "041050",
    year = "2021"
}

@article{TessmerGopakumar2006,
    author = "Tessmer, Manuel and Gopakumar, Achamveedu",
    title = "{Accurate and efficient gravitational waveforms for certain galactic compact binaries}",
    eprint = "gr-qc/0610139",
    archivePrefix = "arXiv",
    doi = "10.1111/j.1365-2966.2006.11179.x",
    journal = "Mon. Not. Roy. Astron. Soc.",
    volume = "374",
    pages = "721--728",
    year = "2007"
}

@article{TanayHaneyGopakumar2016,
    author = "Tanay, Sashwat and Haney, Maria and Gopakumar, Achamveedu",
    title = "{Frequency and time domain inspiral templates for comparable mass compact binaries in eccentric orbits}",
    eprint = "1602.03081",
    archivePrefix = "arXiv",
    primaryClass = "gr-qc",
    doi = "10.1103/PhysRevD.93.064031",
    journal = "Phys. Rev. D",
    volume = "93",
    number = "6",
    pages = "064031",
    year = "2016"
}

@article{Andersson2022,
    author = "Andersson, Fredrik and Riegger, Franziska and Ferraioli, Luigi and Giardini, Domenico and Robertsson, Johan",
    title = "{Tri-linear representations for the Laser Interferometer Space Antenna}",
    doi = "10.1209/0295-5075/ac949a",
    journal = "EPL",
    volume = "140",
    number = "1",
    pages = "19001",
    year = "2022"
}

@article{Riegger2024,
    author = "Riegger, Franziska and Robertsson, Johan and Andersson, Fredrik",
    title = "{Fast and accurate method to simulate the LISA response to Galactic binaries}",
    doi = "10.1103/PhysRevD.110.082001",
    journal = "Phys. Rev. D",
    volume = "110",
    number = "8",
    pages = "082001",
    year = "2024"
}

@misc{LISAConventions2026,
    author = "Baghi, Quentin and others",
    title = "{LISA science ground segment conventions}",
    eprint = "2603.22377",
    archivePrefix = "arXiv",
    primaryClass = "astro-ph.IM",
    reportNumber = "LISA-DDPC-SEG-TN-007",
    month = "3",
    year = "2026"
}

@article{RobsonCornishLiu2019,
    author = "Robson, Travis and Cornish, Neil J. and Liu, Chang",
    title = "{The construction and use of LISA sensitivity curves}",
    eprint = "1803.01944",
    archivePrefix = "arXiv",
    primaryClass = "astro-ph.HE",
    doi = "10.1088/1361-6382/ab1101",
    journal = "Class. Quant. Grav.",
    volume = "36",
    number = "10",
    pages = "105011",
    year = "2019"
}

@article{Willems2007,
    author = "Willems, B. and Kalogera, V. and Vecchio, A. and Ivanova, N. and Rasio, F. A. and Fregeau, J. M. and Belczynski, K.",
    title = "{Eccentric double white dwarfs as LISA sources in globular clusters}",
    eprint = "0705.4287",
    archivePrefix = "arXiv",
    primaryClass = "astro-ph",
    doi = "10.1086/521049",
    journal = "Astrophys. J. Lett.",
    volume = "665",
    pages = "L59",
    year = "2007"
}

@article{Korol2024NSWD,
    author = "Korol, Valeriya and Igoshev, Andrei P. and Toonen, Silvia and Karnesis, Nikolaos and Moore, Christopher J. and Finch, Eliot and Klein, Antoine",
    title = "{Neutron star {\textendash} white dwarf binaries: probing formation pathways and natal kicks with LISA}",
    eprint = "2310.06559",
    archivePrefix = "arXiv",
    primaryClass = "astro-ph.HE",
    doi = "10.1093/mnras/stae889",
    journal = "Mon. Not. Roy. Astron. Soc.",
    volume = "530",
    number = "1",
    pages = "844--860",
    year = "2024"
}

@article{Moore2024EccentricNS,
    author = "Moore, Christopher J. and Finch, Eliot and Klein, Antoine and Korol, Valeriya and Pham, Nhat and Robins, Daniel",
    title = "{Discovering neutron stars with LISA via measurements of orbital eccentricity in galactic binaries}",
    eprint = "2310.06568",
    archivePrefix = "arXiv",
    primaryClass = "astro-ph.HE",
    doi = "10.1093/mnras/stae1288",
    journal = "Mon. Not. Roy. Astron. Soc.",
    volume = "531",
    number = "2",
    pages = "2817--2829",
    year = "2024"
}

@article{Rajamuthukumar2025,
    author = {Rajamuthukumar, Abinaya Swaruba and Korol, Valeriya and Stegmann, Jakob and Preece, Holly and Pakmor, R{\"u}diger and Justham, Stephen and Toonen, Silvia and de Mink, Selma E.},
    title = "{The role of triple evolution in the formation of LISA double white dwarfs}",
    eprint = "2502.09607",
    archivePrefix = "arXiv",
    primaryClass = "astro-ph.SR",
    doi = "10.1051/0004-6361/202554277",
    journal = "Astron. Astrophys.",
    volume = "704",
    pages = "A156",
    year = "2025"
}

@article{Hellstrom2025,
    author = {Hellstr{\"o}m, Lucas and Giersz, Miros{\l}aw and Askar, Abbas and Hypki, Arkadiusz and Zhao, Yuetong and Lu, Youjun and Zhang, Siqi and V{\'a}zquez-Aceves, Ver{\'o}nica and Wiktorowicz, Grzegorz},
    title = "{Formation channels of gravitationally resolvable double white dwarf binaries inside globular clusters}",
    eprint = "2506.13122",
    archivePrefix = "arXiv",
    primaryClass = "astro-ph.SR",
    reportNumber = "702",
    doi = "10.1051/0004-6361/202555960",
    journal = "Astron. Astrophys.",
    volume = "702",
    pages = "A131",
    year = "2025"
}

@article{Karnesis2021,
    author = "Karnesis, Nikolaos and Babak, Stanislav and Pieroni, Mauro and Cornish, Neil and Littenberg, Tyson",
    title = "{Characterization of the stochastic signal originating from compact binary populations as measured by LISA}",
    eprint = "2103.14598",
    archivePrefix = "arXiv",
    primaryClass = "astro-ph.IM",
    doi = "10.1103/PhysRevD.104.043019",
    journal = "Phys. Rev. D",
    volume = "104",
    number = "4",
    pages = "043019",
    year = "2021"
}

@article{DGI,
    author = "Damour, Thibault and Gopakumar, Achamveedu and Iyer, Bala R.",
    title = "{Phasing of gravitational waves from inspiralling eccentric binaries}",
    eprint = "gr-qc/0404128",
    archivePrefix = "arXiv",
    doi = "10.1103/PhysRevD.70.064028",
    journal = "Phys. Rev. D",
    volume = "70",
    pages = "064028",
    year = "2004"
}

@misc{FastGB,
    title = "{{fastGB} package}",
    howpublished = "{\url{https://lisa.pages.in2p3.fr/fastgb/README.html}}",
    year = "2026",
    note = "{Accessed 2026-08-23}"
}

@misc{JAXGB,
    author = "Bayle, Jean-Baptiste and Le Jeune, Maude and Menu, Jacques",
    title = "{{JAXGB}: Fast {LISA} response for Galactic binaries using {JAX}}",
    publisher = "Zenodo",
    year = "2026",
    doi = "10.5281/zenodo.18343519"
}

@misc{PyTDI,
    title = "{{pyTDI} package}",
    howpublished = "{\url{https://pypi.org/project/pyTDI/}}",
    year = "2026",
    note = "{Accessed 2026-08-23}"
}

@article{DD85,
    author = "Damour, Thibault and Deruelle, Nathalie",
    title = "{General relativistic celestial mechanics of binary systems. {I}. The post-newtonian motion}",
    journal = "Annales de l'I.H.P. Physique th{\'e}orique",
    volume = "43",
    pages = "107--132",
    year = "1985",
    url = "https://www.numdam.org/item/AIHPA_1985__43_1_107_0/"
}

@article{DD86,
    author = "Damour, Thibault and Deruelle, Nathalie",
    title = "{General relativistic celestial mechanics of binary systems. {II}. The post-newtonian timing formula}",
    journal = "Annales de l'I.H.P. Physique th{\'e}orique",
    volume = "44",
    pages = "263--292",
    year = "1986",
    url = "https://www.numdam.org/item/AIHPA_1986__44_3_263_0/"
}

@article{Mikkola1987,
    author = "Mikkola, S.",
    title = "{A cubic approximation for Kepler's equation}",
    journal = "Celestial Mechanics",
    volume = "40",
    pages = "329--334",
    year = "1987",
    doi = "10.1007/BF01235850"
}

\end{document}